# Molecular rules for selectivity in lipase-catalyzed acylation of lysine.

Dettori L.[1] Jelsch C. [2] Guiavarc'h Y. [1] Delaunay S. [1] Framboisier X. [1] Chevalot I.[1] and Humeau C. [1]

1 Laboratoire Réactions et Génie des Procédés, CNRS UMR-7274, Université de Lorraine, 2 av. de la Forêt d'Haye, 54500 Vandoeuvre-lès-Nancy, France

2 Laboratoire de Cristallographie, Résonance Magnétique et Modélisations, Institut Jean Barriol, CNRS-UMR 7036, Université de Lorraine, Vandoeuvre-lès-Nancy, France

* catherine.humeau@univ-lorraine.fr

Graphical abstract

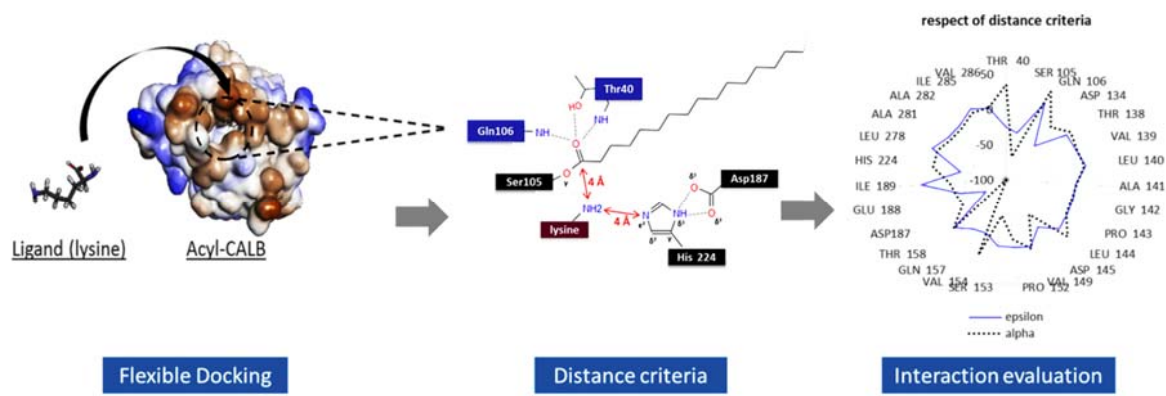

Highlights:

- Methodology development for predicting feasability CALB toward N-acylation.
- Methodology development for predicting selectivity of CALB toward N-acylation.
- Models predicted the N-acylation of L-lysine with lauric acid.

**Abstract**

The selectivity of L-lysine acylation by lauric acid catalysed by *Candida antarctica* lipase B (CALB) was investigated combining experimental and theoretical methodologies. Experiments showed the near-exclusive acylation of lysine ε-amino group; only traces of product resulting from the acylation of lysine α-amino group were observed fleetingly. Molecular modelling simulations were performed aiming to understand the molecular rules for selectivity. Flexible docking simulations combined with structural investigations into lysine/CALB binding modes also suggested the preferential acylation of lysine ε-amino group without, however, excluding the acylation of the lysine α-amino group. Electrostatic interaction energy between lysine and the residues covering the catalytic cavity was calculated in order to understand the discrimination between the two lysine amino groups. The results suggests that the proximity of the carboxylate group hinders the binding of the substrate in configurations enabling the Nα-acylation. Key interactions with the polar region covering the catalytic triad were identified and a plausible explanation for selectivity was proposed.

1. Introduction

Amino acids and peptides include molecules presenting various positive effects (antioxidant, opioids, anti-hypertensive, anti-cancer...) that are receiving increasing interest from pharmaceutical, cosmetic and nutraceutical industries (Chi et al., 2015; Harris et al., 2013; Hernández-Ledesma et al., 2011; Lindqvist et al., 2015). However, their polarity may limit their transfer through cell membranes and their efficiency may be restricted by their short half-life due to potential hydrolysis catalysed by endo-proteases. One possibility to circumvent these problems consists in acylating their structure with a fatty acid chain, leading to derivatives with improved activities and/or new techno-functional properties (Bordes and Holmberg, 2015; Craik et al., 2013). Previous studies showed that acylation of peptides could facilitate their transport through biological membranes by increasing their lipophilic properties. In addition, this could protect them against endogenous proteolytic attack and thus increase their stability (Nestor and others, 2009). Acylated derivatives of amino acids are commonly used as surfactants that are important ingredients in cosmetics, due to their excellent surface-active properties, combined with interesting biological activities and low potential for toxicity (Morán et al., 2004). Acylation process can be performed by either enzymatic or chemical pathways. Major advantages of using enzymes instead of chemical catalysts are a high selectivity leading to the grafting of acyl chains on specific positions and the possibility to apply mild reaction conditions that respect the green chemistry requirements. Hydrolytic enzymes and mainly lipases (EC 3.1.1.3) are most often used to catalyse such reactions in non-aqueous media, and besides already applied to industrial bioprocesses in the food, pharmaceuticals and cosmetics area (Nestl et al., 2011). In addition to broad substrate specificities, lipases do not require any coenzyme. Moreover, some of them are highly stable, especially when immobilized. The lipase B from *Candida Antarctica* (CALB) has already been used in many acylation bioprocesses applied to a wide range of compounds as vitamins, sugars, amino acids or phenols (Ganske and Bornscheuer, 2005; Husson et al., 2011; Soo et al., 2004). Development of such bioprocesses and control of selectivity requires that enzyme/substrates binding modes as well as molecular rules for selectivity are understood in depth. To achieve this, several studies reported the use of molecular modelling simulations, especially based on docking methodologies.

In case the enzyme structure is known, docking appears as an efficient method to predict the preferred orientations of substrates within the target catalytic cavity and then to analyze interactions (Leach et al., 2006; Yuriev and Ramsland, 2013). Poses issued from docking calculations are ranked using score functions, based on binding affinity approximation (Gohlke and Klebe, 2002). This ranking step represents a decisive stage in docking protocols that may strongly impact the accuracy of the models. Many molecular modelling studies aiming to understand substrate specificity as well as selectivity have already been performed on CALB-catalyzed reactions. An approach combining docking and molecular dynamics simulations was applied to study the regioselectivity of CALB during flavonoid acetylation at the molecular level (Bidouil, 2012; Oliveira, 2009). Similarly, the acylation of lysine-based peptides was studied and the regioselectivity of the reaction was successfully predicted (Ferrari et al., 2014). Regardless of the peptide structure, simulation results suggested the preferential N-acylation of the lysine lateral chain. This was also evidenced in experiments. The lysine structure has two amino groups that are prone to acylation. One is located at the end of the lysine lateral chain (position ε), and the other on the carbon α. Lysine acylation catalyzed by the lipase from *Rhizomucor miehei* has already been investigated experimentally (Montet et al., 1990; Soo et al., 2004). Results showed the exclusive acylation of lysine ε-amino group but no clear explanation was given for this selectivity. The present work aimed to study CALB-catalyzed acylation of lysine with lauric acid. Experimental evidence was given for selectivity whereas models brought a deep knowledge of lysine/CALB binding modes. Molecular rules for selectivity were determined through electrostatic interaction energy calculations.

## 2. Materials and methods

### 2.1. Enzymatic reactions

#### 2.1.1. Chemicals and enzyme

Novozym 435® (lipase B from *C. antarctica* immobilized on an acrylic resin) with propyl laurate synthesis activity of 7000 PLU g$^{-1}$ and protein grade of 1–10% was from Novo Nordisk A/S (Bagsværd, Denmark). Lysine and lauric acid were purchased from Sigma-Aldrich (Saint Quentin Fallavier, France). 2-Methyl-2-butanol (M$_2$B$_2$) and triethylamine of HPLC quality were purchased from Carlo Erba (France).

#### 2.1.2. Syntheses

L-lysine (0,12 M) and Lauric acid (0,24 M) were solubilized in 2 mL of 2-methyl-2-butanol previously dehydrated on 4 Å molecular sieves, for 12 h at 55°C. Amines have increased pKa values in organic solvents compared to aqueous solutions (Rõõm et al., 2007). To ensure that the amine groups of the lysine were not protonated a large excess of triethylamine was added to the reaction medium (2.4 mol L$^{-1}$). Only scarce acylation reaction did actually occur when no triethylamine was added to the medium (results not shown). The acylation reaction was started by the addition of 10 g.L$^{-1}$ of the enzymatic preparation Novozym 435®. The reaction medium was stirred at 250 rpm and kept at 55°C. Samples were withdrawn over time, diluted with methanol/water (80/20, v/v), and then, stored at ambient temperature before LC–MS analysis. At the end of the reaction, the enzyme was removed by filtration. Each reaction was repeated at least thrice.

#### 2.1.3. Analysis of reaction media

The synthesis of lysine derivatives was followed on a HPLC–MS-MS system (ThermoFisher Scientific, San Jose, CA, USA) consisting in a binary delivery pump connected to a photodiode array detector (PDA) and a LTQ ion trap as mass analyzer (Linear Trap Quadrupole) equipped with an atmospheric pressure ionization interface operating in positive electrospray mode (ESI$^+$). Chromatographic separation was performed on a C18 column (150 mm × 2.1 mm, 5 μm

porosity – Grace/Alltech, Darmstadt, Germany) equipped with a C18 pre-column (7.5 mm × 2.1 mm, 5 μm porosity – Grace/Alltech Darmstadt, Germany) at 25°C. Mobile phases consisted in methanol/water/TFA (80:20:0.1, v/v/v) for the phase A and methanol/TFA (100:0.1, v/v) for the phase B. Acylated peptides were eluted using a linear gradient from 0% to 100% of B for 5 min and then an isocratic step at 100% of B for 10 min, at a flow rate of 0.2 mL min$^{-1}$. Mass spectrometric conditions were as follows: spray voltage was set at +4.5kV; source gases were set for sheath gas, auxiliary gas and sweep gas at 30, 10 and 10, respectively (in arbitrary units min$^{-1}$); capillary temperature was set at 250°C; capillary voltage was set at 48 V; tube lens, split lens and front lens voltages were set at 120 V, -34 V and -4.25 V, respectively. Full scan MS spectra were performed from 100 to 1000 m/z and additional MS$^2$ scans were realized in order to get structural information based on daughter ions elucidation. Raw data were processed using Xcalibur software (version 2.1 Thermo Scientific).

### 2.2. Molecular modeling simulations

#### 2.2.1. Computational resources

Several modules of the software Discovery Studio 4.1 (Accelrys, Inc.) were used in this study:

- CHARMm: force field for molecular mechanics (Brooks et al., 1983)
- Flexible Docking: module for docking simulations including receptor and ligand flexibility (Koska et al., 2008)
- Calculate binding energies: module for estimation of binding energy between a receptor and a ligand (Tirado-Rives and Jorgensen, 2006)

Simulations were carried out on a computer cluster equipped with 2 Quad Core Intel Xeon processors L5420, 2.5 GHz and 16 GB of RAM on a Linux Platform.

#### 2.2.2. Lysine construction

The three-dimensional structure of lysine was built using the Builder module of the program-package Discovery Studio 4.1. All atoms were typed by applying the general CHARMm force field. The resulting structure was submitted to a geometric correction aiming to remove atomic disorder.

### 2.2.3. Acyl-enzyme construction

The 1LBS PDB entry was chosen as the reference structure for CALB. The 1LBS crystal structure is composed of six independent protein chains per asymmetric unit. Each of these chains consists of 317 amino acids, one ethyl-hexyl-phosphonate (HEE) inhibitor covalently bound to the catalytic serine (Ser105), one N-acetyl-glucosamine (NAG) dimer covalently bound to the residues Asn74. The main interest of this structure is the presence of the inhibitor that provides experimental evidence for the orientation of the acyl moiety towards the oxyanion hole and more generally, towards the residues constituting the catalytic cavity. The acyl-enzyme model was built according to the procedure previously described by Ferrari *et al*. (2014) with adaptation to lauroyl chain (Ferrari et al., 2014). Asp134 was modelled as protonated as observed in high resolution crystal structure (Stauch et al., 2015).

### 2.2.4. Flexible Docking protocol

A protocol of docking including the flexibility of both the ligand and the lateral chain of residues constituting the binding site was used (Koska et al., 2008). The first step used the ChiFlex algorithm (CHARmm based molecular mechanics) to explore the conformational space of flexible residues and generate a set of low energy conformations of the protein (Spassov et al., 2007). For each of these, hotspots corresponding to polar and non-polar zones were identified. Parallely, the Catconf algorithm was used to search low energy conformations of the ligand (~50 conformers.mn$^{-1}$) (Li et al., 2007; Smellie et al., 1995a, 1995b, 1995c). Then, each ligand conformation was docked rigidly against each conformation of the protein aligning with the hotspots, using the LibDock program (Diller and Merz, 2001). Finally, refinement of side chain positions in the presence of the ligand and then, final refinement of docked poses in the presence of the receptor were performed with ChiRotor and CDOCKER, respectively (Wu et al., 2003). For each final pose, the CHARMm energy and the interaction energy were calculated. The poses were sorted by the CHARMm energy and the top scoring poses were retained.

In the present study, the flexibility was applied to the residues: Thr40, Ser105, Gln106, Asp134, Leu140, Ala141, Leu144, Val149, Val154, Gln157, Asp187, Ile189, Hist224, Leu278, Ala281,

Ala282, Ile285 and Val286 (**Erreur ! Source du renvoi introuvable.**). The root mean square deviation (RMSD) was computed on these residues, referring to the input structure.

### 2.2.5. Analysis of docking results

#### 2.2.5.1. Distance criteria

Inside the active site of CALB, five residues play an important role in the catalytic mechanism: Ser105, His224 and Asp187 residues forming the catalytic triad and the oxyanion hole residues Gln106 and Thr40. The catalytic serine is located at the bottom of the cavity described as a deep and narrow channel (10Å x 4Å wide and 12Å deep) (Uppenberg et al., 1995, 1994). The two oxyanion hole residues stabilize the reaction intermediates through hydrogen bond interactions with the main chain N-H moieties and Thr40 O-H hydroxyl. It was also observed that the active site cavity can be separated into two areas, one of which is globally hydrophobic and the other mostly hydrophilic. The acyl donor substrate is assumed to lodge all along the hydrophobic area whereas the acyl acceptor substrate is acknowledged to access the catalytic triad through the hydrophilic area (Pleiss et al., 1998). The residues Asp134, Gln106, and Thr40 constitute a polar region surrounding the catalytic serine. Based on these structural insights, the likeliest complexes issued from docking simulations were analysed taking into account two distance requirements (**Erreur ! Source du renvoi introuvable.**):

- → [1]: the nitrogen atom to be acylated must be at a distance lower than 4 Å from the acyl-enzyme electrophilic center so that a nucleophilic attack can happen.
- → [2]: the acidic proton of the amino group to be acylated must be at a distance lower than 4 Å from the catalytic histidine so that a proton transfer could occur.

Respect or non-respect of these criteria has previously (De Oliveira et al., 2009) been presumed to correlate with the formation of productive complexes that could lead to the formation of the 2$^{nd}$ tetrahedral intermediate and then to the product.

#### 2.2.5.2. Electrostatic interaction energy

To obtain accurate electrostatic energies, a multipolar atom model was used. The electron density was transferred from ELMAM2 database modelling protein atom types and common chemical groups (Domagała et al., 2011). The interaction energy values were computed as an integral over the electron density $\rho_A$ of molecule A multiplied by the electrostatic potential $\varphi_B$ of molecule B, or reciprocally (Domagała et al., 2011).

$$E_{elec} = \int \rho_A \varphi_B dr_A = \int \rho_B \varphi_A dr_B$$

Determination of the electrostatic interaction energy was executed by using numerical method of exact integration around selected atoms (Volkov et al., 2004), taking into account the electron density overlap of atoms, as implemented in the MoProSuite software (Jelsch et al., 2005). The methodology was for instance applied to complexes of human aldose reductase to discriminate the electrostatic energy of ligand inhibitors (Fournier et al., 2009).

## 3. Results and discussion

### 3.1. Study of lysine binding modes within CALB

#### 3.1.1. Flexible Docking results

A molecular modelling approach was developed to study the regioselectivity of lysine acylation catalysed by CALB, assuming that the binding mode of the amino acid within the catalytic cavity plays a decisive role. The strategy consisted in building a target-model of acyl-enzyme firstly and then, to submit lysine to flexible docking simulations. 41 complexes, also called "poses", associating the acyl-enzyme lauroyl-CALB and the acyl acceptor lysine were obtained. These poses were grouped according to the orientation of lysine within the cavity, either its α- or its ε-amino group pointing towards the catalytic triad. For more clarity, these two orientations are designated by the symbols α and ε in the remainder of the article. 26 out of the 41 poses showed the lysine α-amino group orientated towards the cavity bottom (α orientation), whereas the 15 other poses showed a reversed orientation of the lysine (ε orientation).

These complexes were then subjected to various supplementary analyses aiming to deepen knowledge and understanding of lysine binding modes. First, the active site flexibility was studied. The choice of the flexible residues fell on the catalytic triad Ser105, His224 and Asp187, the oxyanion hole residues Thr40 and Gln106, and several residues covering the hydrophobic area of the cavity Leu140, Ala141, Leu144, Val149, Ser150, Ala151, Val154, Ile189, Lys290, Leu278, Ala281, Ala282, Ile285 and Ala286. The root mean square deviation (RMSD) applied to the side-chain of flexible residues was calculated by comparing with the input acyl-enzyme structure. Whatever the orientation α or ε of the lysine ligand, the most flexible residues were Ile189, Asp187, His224, Asp134, Gln106, with a maximum for Gln157 (RMSD$_α$ = 2.00±0.08 Å and RMSD$_ε$ = 2.00±0.21 Å). The orientation α led to a higher RMSD value for the residue His224 than the orientation ε (RMSD$_α$ = 0.94±0.12 Å and RMSD$_ε$ = 0.48±0.21 Å). Conversely, the residue Asp187 was more mobile in the orientation ε compared with the orientation α (RMSD$_α$ = 0.68±0.14 and RMSD$_ε$ = 1.08±0.24). However, this residue is not directly in contact with the ligand.

Figure shows the two main orientations of the His224 lateral chain observed among the complexes. Important distances ensuring that the active conformation of the enzyme is maintained and that the catalytic mechanism could occur are also indicated.

The conformation presented on Figure .A respected the distances allowing hydrogen atom transfers involved in the catalytic pathway. In contrast, in some complexes, the His224 lateral chain was observed to rotate leading to a wrong conformation and then distance to Asp187 was increased (Figure .B). The capacity of the poses issued from docking to react was studied through two distance measurements (**Erreur ! Source du renvoi introuvable.**). The first one corresponded to the distance between the electrophilic carbon atom of the acyl donor (lauroyl:C) and the nucleophilic nitrogen atom of the acyl acceptor (lysine:N) (distance 1). The second one provided information on the proximity between the hydrogen atom of the group becoming acylated and the His224:Nε (distance2). These distance criteria have already been successfully used to explain and even predict the selectivity of acylation reactions (De Oliveira et al., 2009; Ferrari et al., 2014). When the lysine was in the α orientation, the distance 1 was lower than 4Å for 8 out of 26 poses, vs. 7 poses out of 15 in the case of the ε orientation. These results showed that, whatever the orientation of the lysine, the substrate could approach the electrophilic centre of the acyl-enzyme. The distance 2 seemed more difficult to meet and was respected in 4 poses out of 26 for the orientation α vs. 5 out of 15 for the orientation ε. Combining the two distance criteria led to 2 and 4 correct poses for the orientations α and ε, respectively. In a previous study, rigid docking simulations were carried out on the peptides SK, SYK, KYS and LQKW against oleoyl-CALB target (Ferrari, 2014). When comparing with the present study, peptides seemed overall closer to the catalytic triad than lysine. However, results remained difficult to compare as both the acyl-enzyme and the ligand were different.

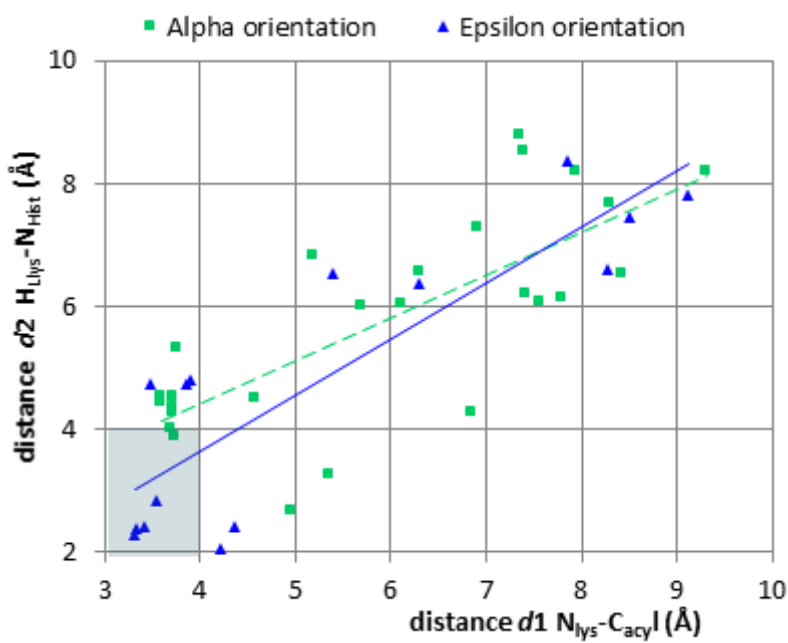

represents the distance $d_1$ according to the distance $d_2$ for each pose issued from docking simulations. Poses were grouped, depending on the orientation α or ε of lysine. Trend lines showed that the distance criteria were respected for some systems belonging to both groups of poses. However, when lysine adopted the α orientation, the proximity of the $NH_2$ moiety to His224 seemed difficult to meet. This result was consistent with previous observations relating to the rotation of the His lateral chain.

Anyway, these theoretical results suggested that the acylation of both amino groups of lysine could occur, leading to two distinct mono-acylated products. A possible limitation to the acylation of the lysine α-amino group could be the hydrogen transfer from the lysine α-amino group to His224, due to the non-favourable orientation of imidazole ring. However, at the docking stage, no theoretical evidence could be found to really discriminate between the two acylation sites.

### 3.1.2. Experimental validation

The synthesis of lauroyl-lysine catalysed by CALB was carried out and the formation of products was monitored through HPLC-MS$^2$ analyses. Acylation reactions were performed at 55°C during 72 h, in 2-methyl-2-butanol as solvent, in the presence of tri-ethylamine (TEA) to control the ionization state of the substrate amino groups (Reyes-Duarte et al., 2002). Lauric acid:lysine molar ratio was 2:1. Mono-lauroyl-lysine products and di-lauroyl-lysine have a molecular mass of 328 ($[M+H]^+$) and 510, respectively. On the basis of mass analysis, the presence of one major mono-acylated product was observed. In order to discriminate between α- and ε-lauroyl-lysine, the corresponding standards were analyzed and their Mass Spectrometry $(MS)^2$ profiles were generated by fragmentation of the parent ions (**Erreur ! Source du renvoi introuvable.**). Fragmentation of α-lauroyl-lysine led to one major daughter ion (m/z = 293) corresponding to $[(Lauroyl-K) – 2H_2O + H]^+$, and two minor daughter ions (m/z = 147 and m/z = 266) corresponding to $[(K) + H]^+$ and $[(lauroyl-K) – H_2O – COOH + H]^+$, respectively. Fragmentation of ε-lauroyl-lysine led to one major daughter ion (m/z = 266) corresponding to $(lauroyl-K) – H_2O – COOH + H]^+$, and one minor daughter ion (m/z = 147) corresponding to $[(K) + H]^+$.

According to these indications, experimental evidence supported that the major product formed during the reaction was ε-lauroyl-lysine. These results are consistent with a previous study about the acylation of various lysine-based peptides catalysed by CALB with oleic acid, showing that the selective acylation of the lysine ε-amino group occurred (Ferrari et al., 2014; Soo et al., 2004). In rare cases, traces of α-lauroyl-lysine were detected at the beginning of some reactions.

Theoretical experiments left a doubt about the selectivity of the reaction although structural elements strongly suggested the preferential acylation of the lysine ε-amino group. It seems that distance criteria required for the catalytic mechanism are more easily met for the ε orientation than for the α orientation but this criterion would not constitute a sufficient basis for bringing theoretical evidence for the selectivity of the reaction. Therefore, energetic criteria were also explored aiming to enhance the robustness of our models and their predictive capacity, while at the same time trying to limit simulation costs.

### 3.2. Energetical analysis of poses

#### 3.2.1. C-DOCKER interaction energy

In order to discriminate poses obtained from the Flexible Docking workflow, the C-DOCKER interaction energy was used (Koska et al., 2008). A correlation research was performed aiming to determine if there was a significant relationship between the distance criteria and the CDOCKER energy. Then, the relevance of these combined parameters to justify the regioselectivity of lysine acylation was evaluated. As shown in Table 1, no significant correlation was found between the CDOCKER energy and the respect of distance criteria. A low 0.33 correlation was found between lysine orientation and the CDOCKER energy, the orientation ε leading to a stronger interaction energy compared with the orientation α. However there was no clear and strong evidence of the selective acylation of the lysine ε-amino group. Overall, it would seem that ranking and post-docking tools, considered separately or combined with distance criteria, were not sufficient to match perfectly with experimental results. Many studies reported that the prediction of binding affinities turns out to be really difficult and require extensive computational time (Irwin and Shoichet, 2016; Leach et al., 2006). Consequently, ranking of poses is identified as a difficult challenge that cannot be ignored when evaluating docking accuracy (Erickson et al., 2004). This question appears increasingly difficult to address when applied to highly flexible ligands or in the case of important conformational changes in the protein structure upon ligand binding.

Table 1: Correlation coefficients between the CDOCKER energy and respect of distance criteria or lysine orientation. MoPro $E_{elec}$ is the sum of electrostatic energies between lysine and the active site residues marked in Fig. 3.4 or between lysine and the residues Ser105 and His224.

| correlation | C-DOCKER Interaction energy | MoPro $E_{elec}$ Electrostatic energy | MoPro $E_{elec}$ E(Ser105)+E(His224) |
|---|---|---|---|
| **Energy / respect of distance criteria** | 0.11 | 0.57 | 0.78 |
| **Energy / lysine orientation** | 0.33 | 0.13 | 0.36 |

### 3.2.2. Electrostatic interaction energy

More detailed research was carried out into the way lysine interacts with CALB catalytic pocket, while maintaining the objectives of discriminating between the two groups of poses (α or ε orientation) and limiting simulation costs. Electrostatic interaction energies using a multipolar atom model were calculated between lysine and the catalytic residues; Figure 7 focuses more specifically on Ser105 and Hist224.

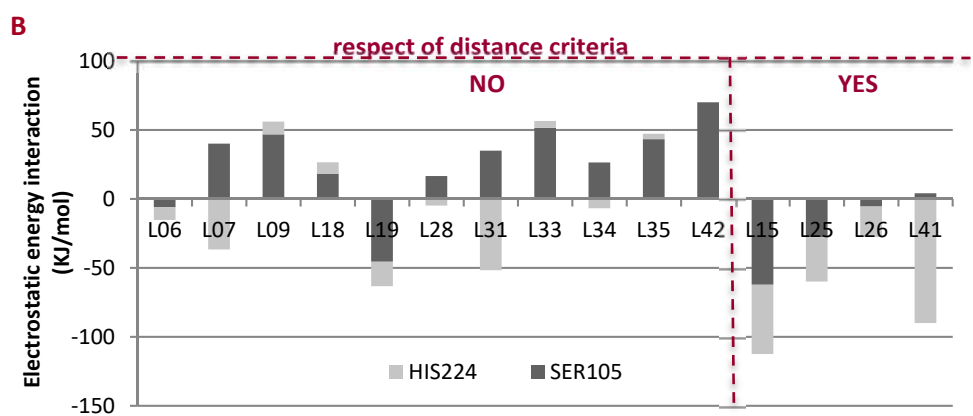

shows electrostatic interaction energies of poses depending on lysine orientation; poses were also grouped depending on their proximity to the catalytic triad. Whatever the lysine orientation, almost all of the poses not respecting the distance criteria led to positive (repulsive) electrostatic interaction energies towards either Ser105 or His224, or more often towards both

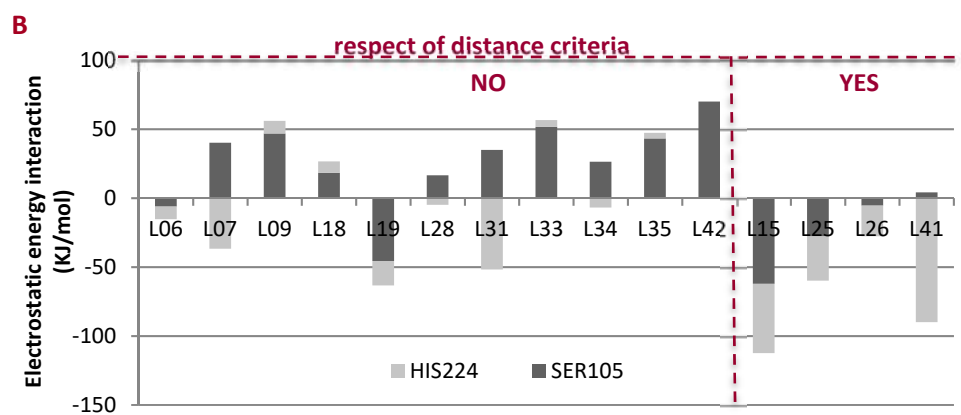

residues (

.A and B). This trend seemed quite understandable as poses not respecting the distance criteria were unlikely to interact in a favorable way with the catalytic residues.

The H-N-H…N$\varepsilon_2$ interaction between one of the lysine NH$_2$ groups and His224 is potentially a hydrogen bond and is electrostatically favorable. For instance, in the $\zeta$ pose L15 complex with strongest $E_{elec}$(Lys,His224)=-50kJ/mol, the hydrogen bond contribution $E_{elec}$(H$\zeta$…N$\varepsilon_2$) reaches -32kJ/mol. On the other hand, the $E_{elec}$(Lys,Ser105Lauroyl)=-62 kJ/mol attractive interaction in complex L15 is mainly due to the hydrogen bond between the lysine N$\zeta$H$_2$ group and the oxygen O=C of the ester group in the acylated Ser105 as $E_{elec}$(H…O)=-50kJ/mol. A few poses not respecting the distance criteria led to negative electrostatic interaction energies towards Ser105 and His224 (L22, L40, L06 and L19), presumably due to attractive interactions between other atoms that were not concerned by the distance measurements.

The two poses in the orientation α respecting the distance criteria led to attractive electrostatic interaction energies towards Ser105 but not towards His224 (

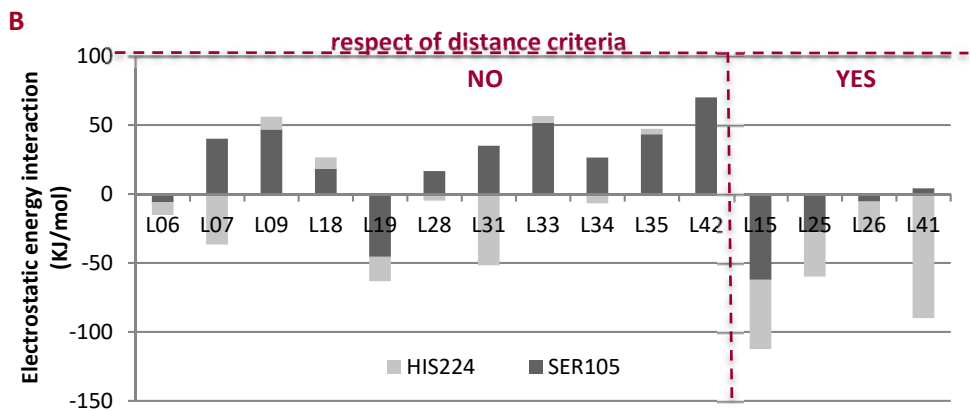

.A).

Inspection of complex L29 shows indeed a relatively long distance $d_2$=3.9Å between the $NH_2$ group and His224. Moreover, the $E_{elec}$ value between lysine and His224 in both productive α complexes is rendered non favorable (Figure 7) due to the proximity of the electronegative lysine carboxylic oxygen atoms with His224: $d(O…Nε_2)$=4.3Å in L29 and 3.3Å in L37 (Figure 6). Interaction diagrams brought out hydrogen bond and van der Waals interactions between the lysine and the residues Ser153, Thr40, Gln157 and Thr40, stabilizing the lysine α-amino-group close to the catalytic serine but far from the catalytic histidine. Poses in the ε orientation respecting the distance criteria gave negative electrostatic interaction energies towards both catalytic residues Ser105 and His224 (

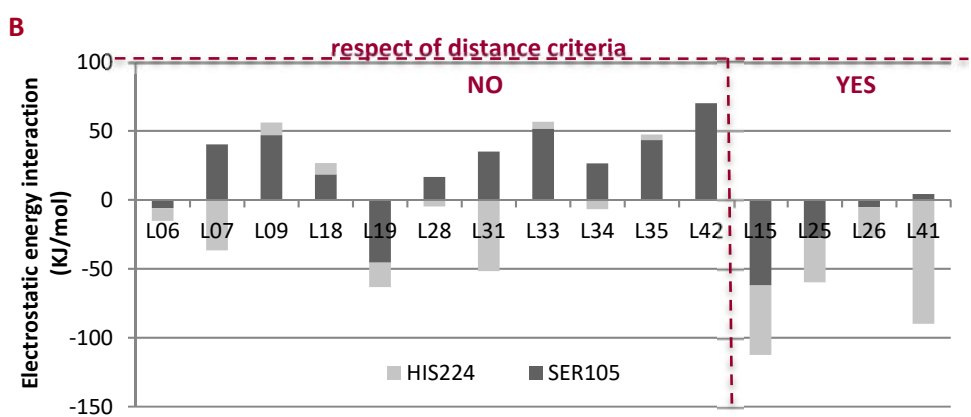

.B: L15, L25, L26). Thus, these poses could be qualified as productive as these were geometrically and energetically favorable to the formation of the tetrahedral intermediate of the resulting reaction product.

These results confirmed that lysine in the ε orientation could successfully approach the catalytic triad and interact with it, contrarily to lysine in the α orientation that came hardly close to the catalytic His224. At this stage, the respect of distance criteria combined with favorable electrostatic interactions with the catalytic residues Ser105 and His224 strongly suggested a trend that perfectly met experimental data, leading to the preferential acylation of the lysine ε-amino group.

The $E_{elec}$ values of lysine with Ser105 and His224 correlate well with the two distance criteria being respected ($c$=78%) which involved these same two residues. The $E_{elec}$ values summed over all the selected active site residues show a correlation of 57% with $d_1+d_2$ criteria. This suggests that ligand poses which are not in catalytic position generally have not recovered electrostatic attractions equivalent to the hydrogen bonds formed with Ser105 and His224.

### 3.3. Molecular rules for selectivity

In order to deepen scientific knowledge of the causes that govern the selectivity in lysine acylation reaction, electrostatic interaction energies between lysine and the residues constituting the catalytic cavity were calculated. Location of these residues is shown on **Erreur ! Source du renvoi introuvable.**. The average $E_{elec}$ energies are presented on **Erreur ! Source du renvoi introuvable.**. Poses were grouped according to their orientation and their proximity to the catalytic triad.

**Erreur ! Source du renvoi introuvable.**.A and B allowed studying the effect of lysine position on electrostatic interaction energies, for one given orientation of the ligand. When lysine adopted the ε orientation close to the cavity bottom, the main interactions were observed with residues of the catalytic triad and the polar Asp134. Interactions with the catalytic Ser105 and His224 logically disappeared in poses not respecting the distance criteria. Thus, the catalytic residues appeared as crucial interaction spots stabilizing lysine in a correct orientation within the cavity. The contribution of these residues to a hydrogen bond network imposing electrostatic interaction restrictions to the substrates was firstly mentioned by researchers who elucidated CALB structure (Uppenberg et al., 1994).

The role of Asp134 in positioning of acyl acceptor substrates was confirmed in a study relative to CALB-catalyzed acylation of flavonoids (Bidouil et al., 2011). In the present study, this residue and lysine substrate are not charged, but have the potential to form strong hydrogen bonds. Indeed the interaction $E_{elec}$ value of lysine with Asp134 was found to be attractive in the ε poses respecting distance criteria in the range -22 to -60 kJ/mol. In the two α poses, the energy was -10 and +5 kJ/mol).

Poses in the orientation α close to the cavity bottom led to interactions with Ser105, Asp187 and other polar residues facing away from the catalytic triad like Gln157, Ser153 and Asp145. The absence of attractive interaction with His224 was coherent with observations relative to distance criteria being generally not fulfilled. Repulsion from the oxyanion hole residues is also suggested. Poses distant from the cavity bottom lost interaction with Ser105 while gaining interaction with Glu188 and Asp187. Whatever the proximity of lysine to the cavity bottom, a lack of interaction with Asp134 was noticed. These results suggested lysine in the α orientation could be stabilized within the cavity but in a wrong configuration that was unlikely to be productive.

**Erreur ! Source du renvoi introuvable.**.C and D compared interactions aiming to determine the driving forces leading to productive or not productive orientation of lysine. In the case of poses far from the catalytic triad, lysine interacted with Asp187, regardless of its orientation. Orientation α led to interaction with Asp145 instead of Asp134 observed for orientation ε. As a reminder, Asp145 is positioned in the active site opposite to the catalytic triad. In the case of poses close to the catalytic triad, differences were observed depending on the orientation of lysine. The orientation ε led to strengthened interactions with the catalytic triad and Asp134. In orientation α, lysine gained interaction with residues facing away from the catalytic triad such as Asp145, Ser153, and Gln157.

## 4. Conclusion

A methodology combining docking simulations and interaction energy calculations was proposed to examine the regioselectivity of lysine acylation catalyzed by *C. antarctica* lipase B at a molecular level. Experiments showed the near-exclusive acylation of lysine ε-amino group; this trend was also suggested by docking simulations. Indeed, the best poses according to the CDOCKER energy revealed two main orientations of the lysine within the catalytic cavity, depending on whether the ε- or the α-amino group of lysine pointed towards the catalytic triad (orientations ε or α, respectively).

However, more detailed analysis of the complexes suggested that the α orientation led to poses where the lysine amino group lies broadly far from the catalytic His224 or where the electrostatic interaction energy between the lysine substrate and His224 is not attractive. Visual inspection of the docked poses shows that in the α orientation, the lysine carboxylate group, which is close to the α-amino group moiety, leads to an electrostatically unfavorable position and may hinder the ligand anchoring. Moreover, the presence of the bulky COOH group near the $NH_2$ group in α position may cause some steric hindrance while the ε position is more accessible and shows more stereochemical/conformational flexibility.

Instead, lysine would seem attracted by other polar residues facing away from the catalytic triad. Binding in the orientation ε appeared quite different, as criteria relative to the proximity of lysine to the catalytic machinery seemed easier to meet compared with the orientation α. Furthermore, favorable interactions were observed between lysine and residues constituting a small polar area covering the catalytic triad and some neighboring residues, namely Asp134.

Calculation of the accurate electrostatic interaction energy using a multipolar atom model appeared as a pertinent method to characterize the interaction network within enzyme/substrates complexes and to discriminate between the models issued from docking. In the present case study, modelling results helped determine the molecular rules for selectivity in lysine acylation. Favorable electrostatic interactions with the catalytic residues His224, Ser105, Asp187 and Asp134 appeared as good indicators of the reaction feasibility. More broadly, global evaluation of electrostatic interactions helped determining the driving forces responsible for the reaction selectivity.

Figures :

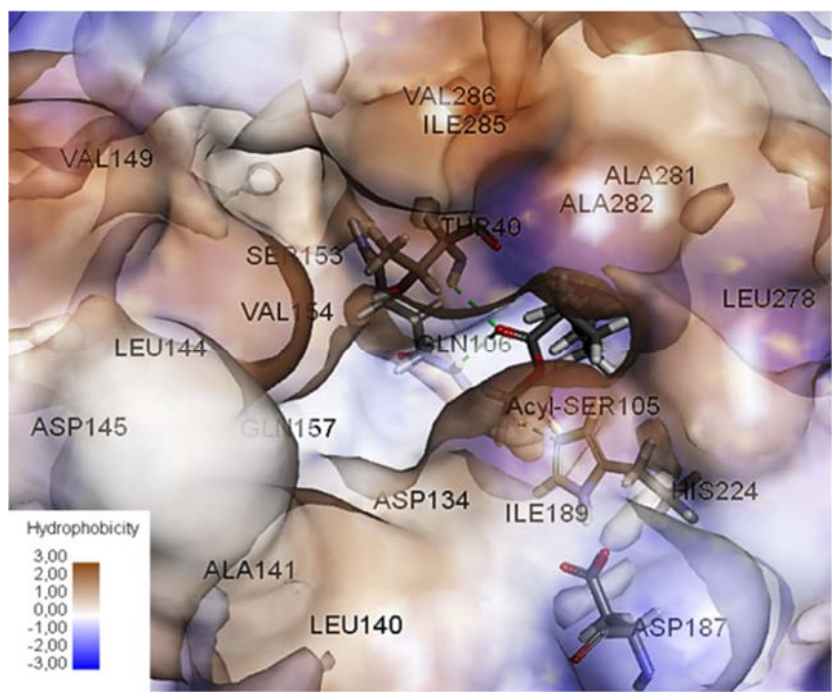

Figure 1: View of the CALB active site pocket. Residues of CALB for which side chains were considered as flexible during docking simulations are marked.

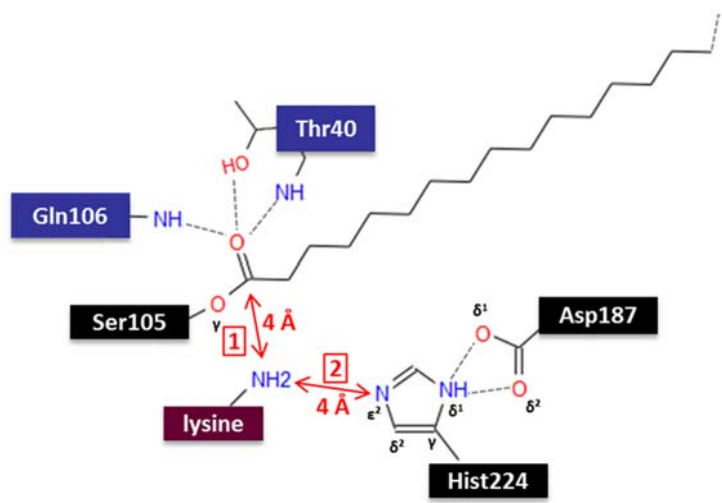

Figure 2: Theoretical distance criteria ($d_1$ and $d_2$) to be met by productive complexes.

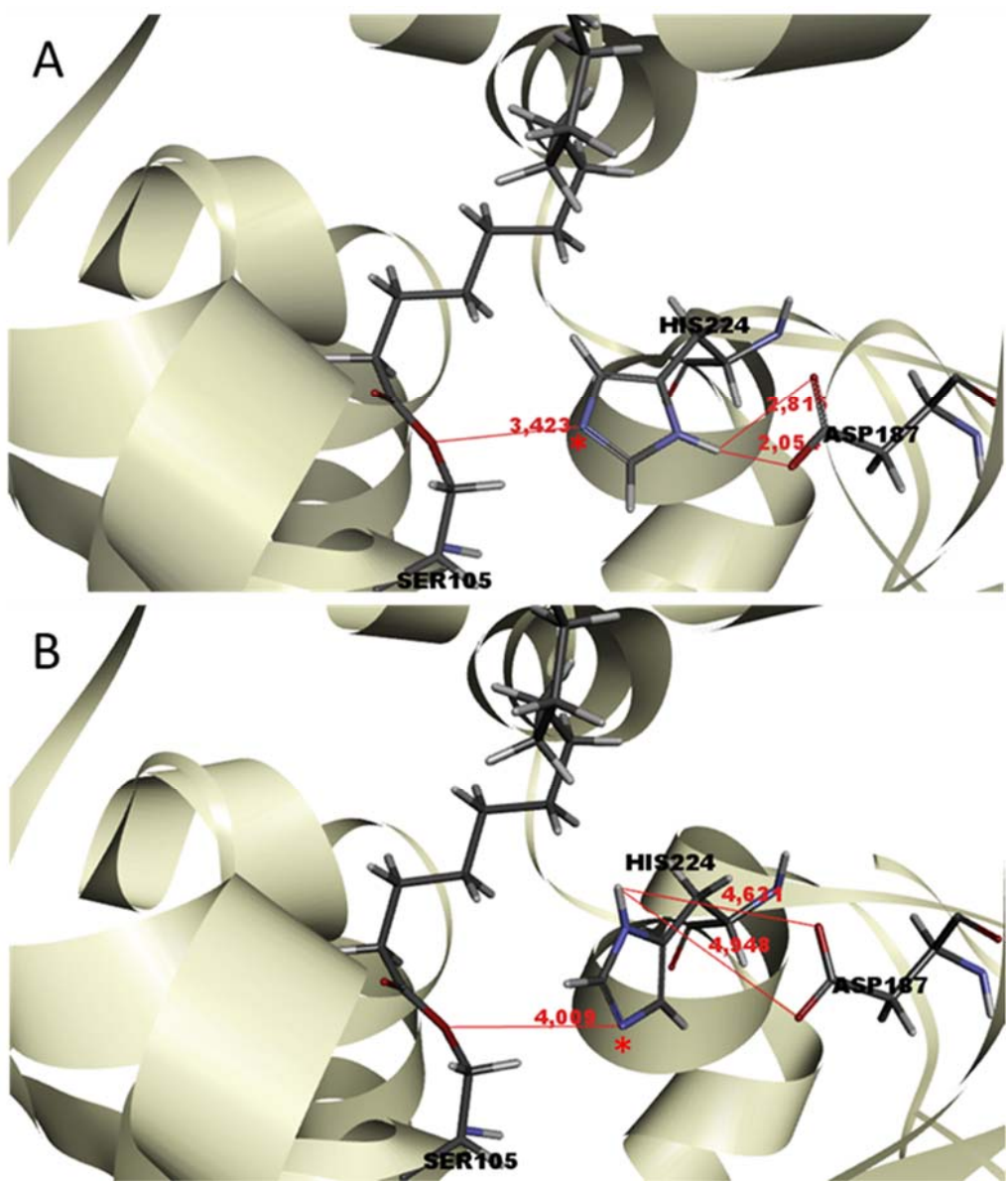

Figure 3

Figure 3: Main conformations of the catalytic cavity observed among the acyl-enzyme complexes issued from docking. A: one example of complex showing a correct conformation of the catalytic machinery. B: one example of complex showing an unfavorable orientation of the His224 lateral chain. Distances theoretically involved in hydrogen atom transfers are indicated (in Å). For more clarity, the lysine is not shown.

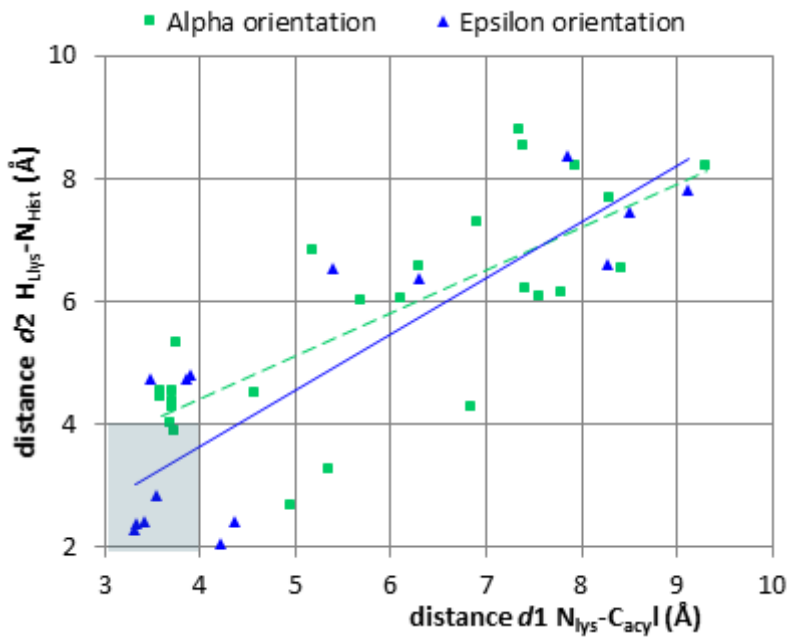

Figure 4: Distances $d_1$ and $d_2$ for each pose issued from docking, depending on lysine orientation. ▲ (–): poses in the orientation ε ; ■ (--): poses in the orientation α.

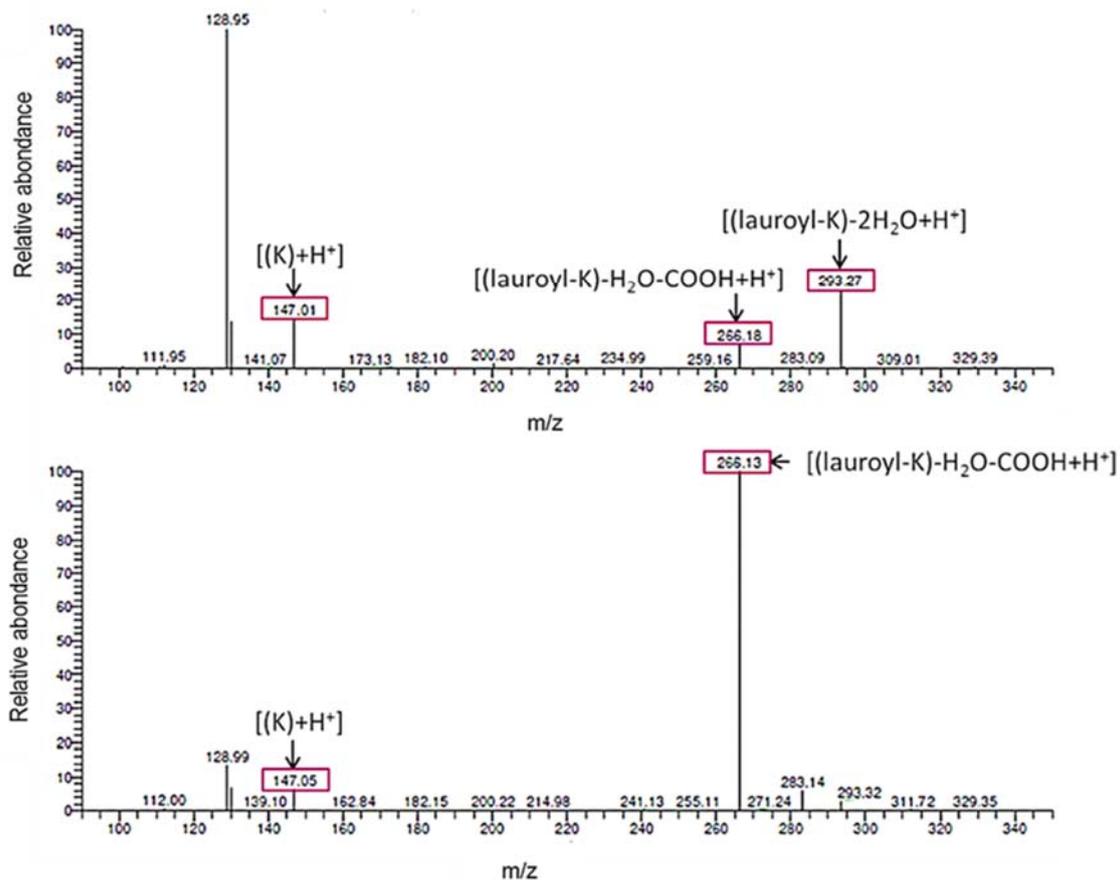

Figure 5: Mass analyses of α/ε-lauroyl-lysine standards (m/z=329). MS$^2$ spectrum after fragmentation of the parent ion.

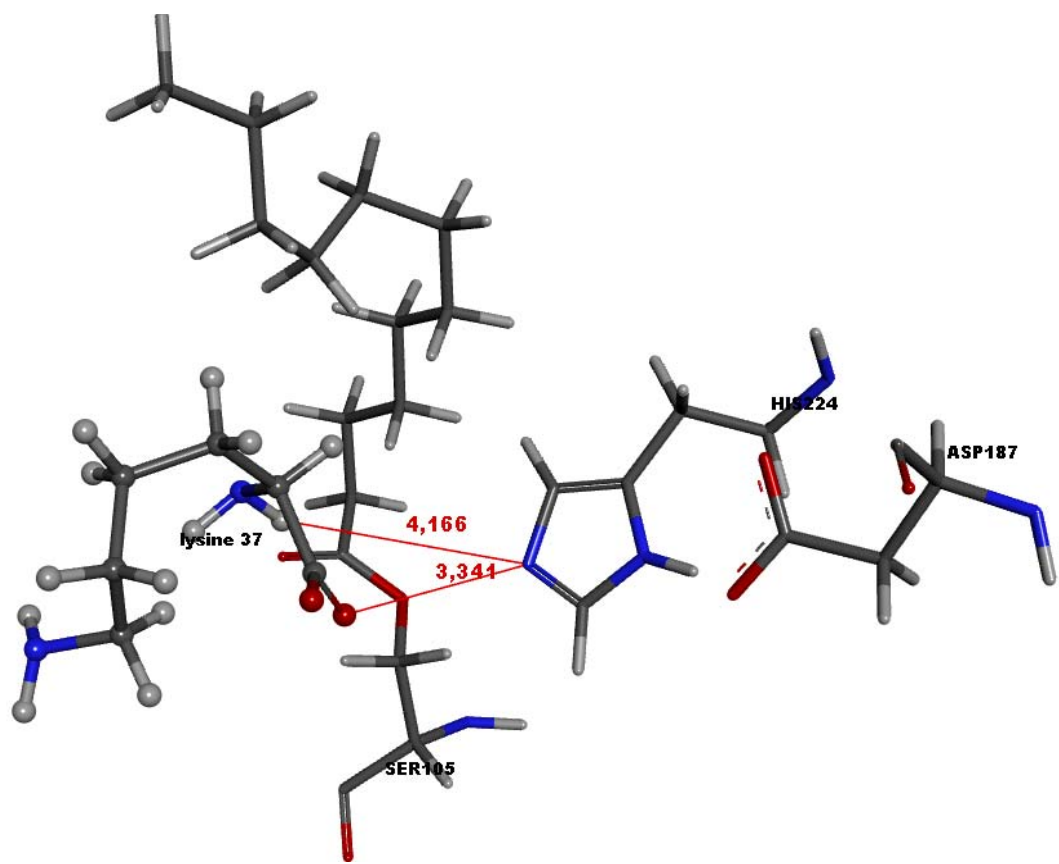

Figure 6: α-Conformation L37 of the catalytic cavity observed among the acyl-enzyme complexes issued from docking.

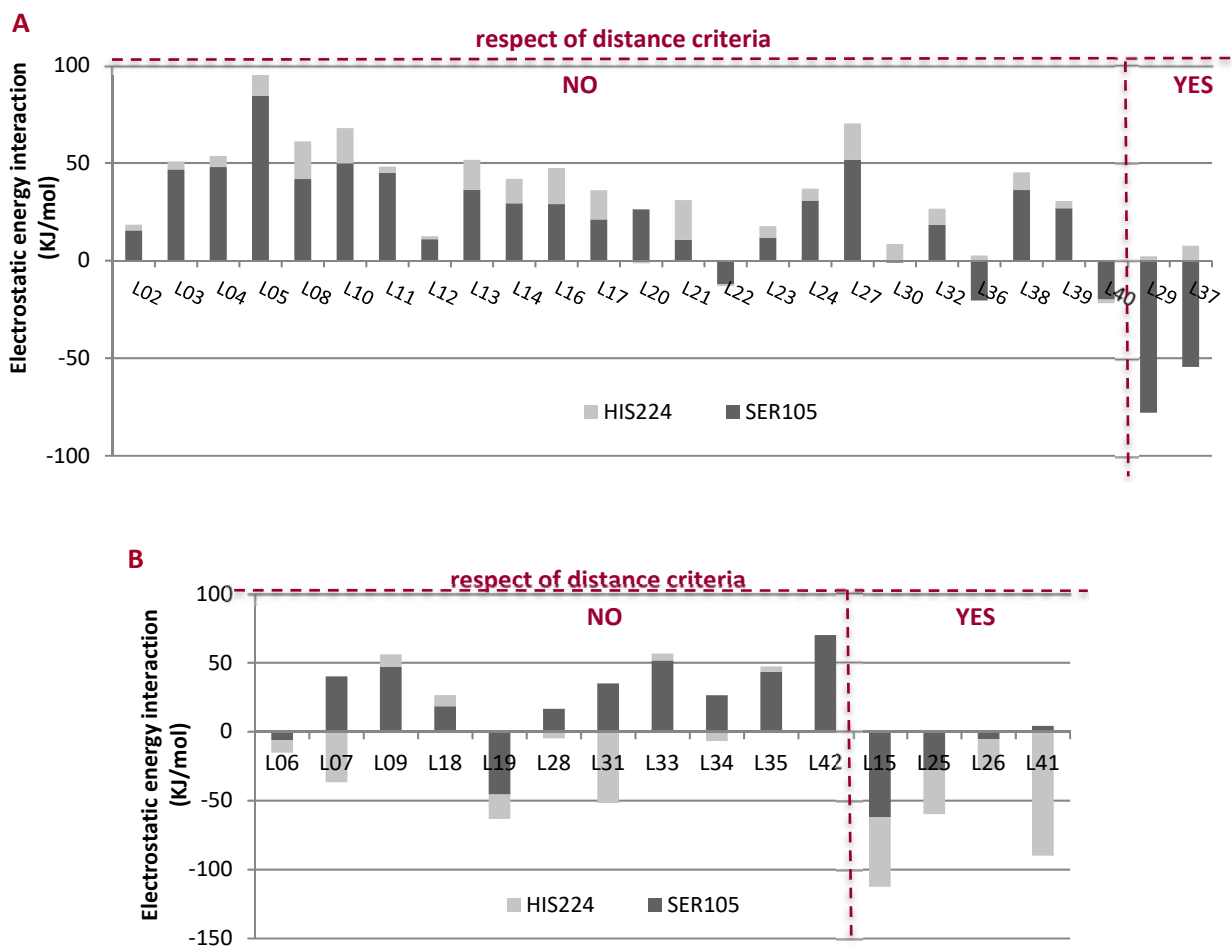

Figure 7: Electrostatic interaction energies (kJ/mol) between the catalytic residues Ser105 and His224 and lysine in the orientation α (A) or ε (B). Poses were grouped according to the proximity of lysine to the catalytic triad. Poses respecting the distance criteria are given on the right.

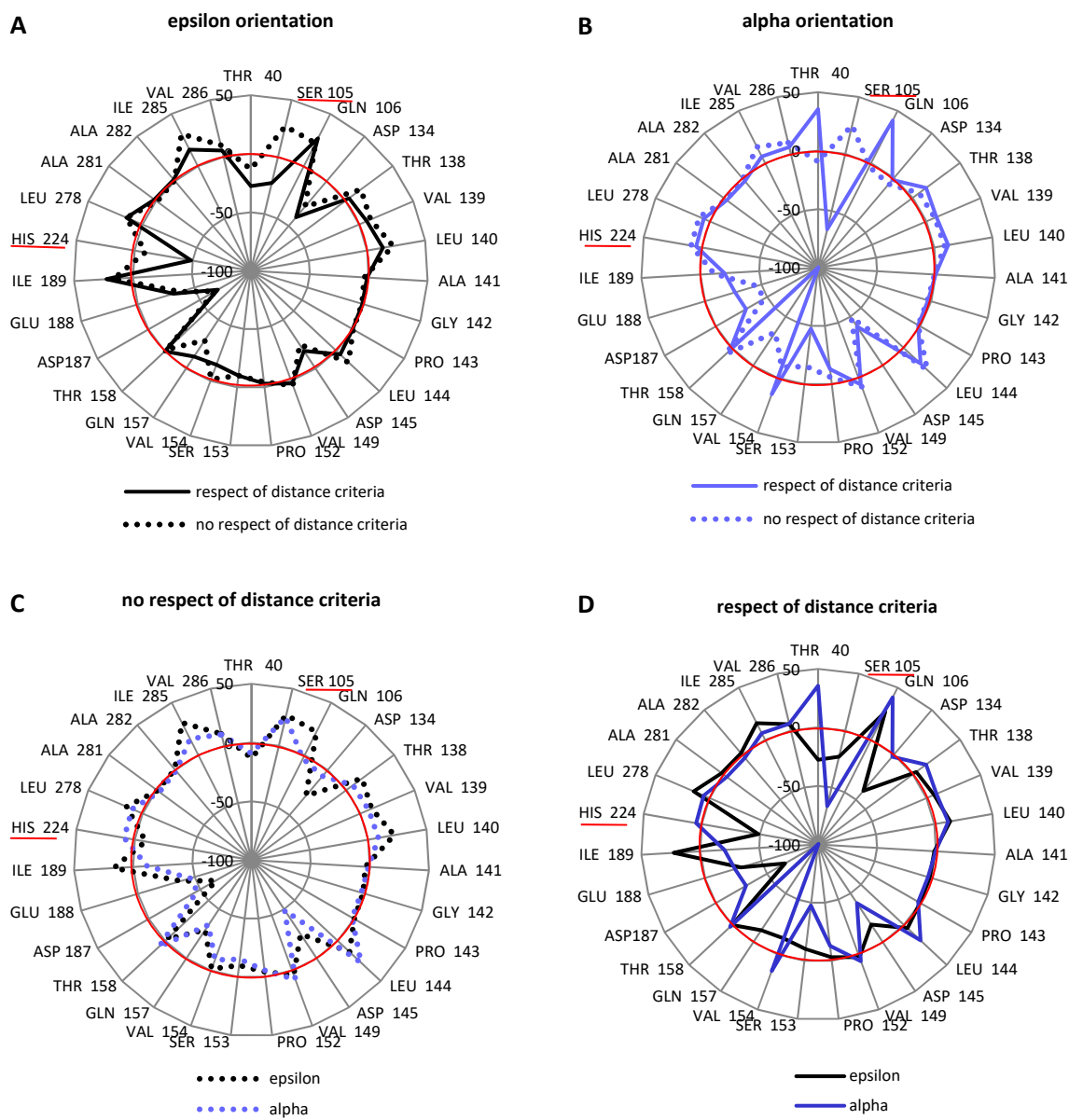

Figure 8: Average electrostatic interaction energies within lysine/acyl-CALB complexes. Poses were grouped according to the orientation of lysine (α in blue or ε in black), and its proximity to the catalytic triad. To ease reference, the red circle represented energy equal to zero.

26